\documentstyle[preprint,aps]{revtex}
\def\be{\begin{equation}}
\def\para{\tau}
\def\ee{\end{equation}}
\def\xf{x_F}
\def\xaf{x_{AF}}
\def\yf{y_F}
\def\yaf{y_{AF}}
\def\trf{\tilde{R}_F}
\def\ttf{\tilde{\theta}_F}
\begin{document}
\draft
\title{\bf
Mean-Field Analysis of Antiferromagnetic
Three-State Potts Model with Next Nearest Neighbor
Interaction}
\author{M.~Itakura}
\address{Department of Pure and Applied Sciences,\\
University of Tokyo,\\
Meguro-ku, Komaba 3-8-1, Tokyo 153, Japan\\}
\date{\today}
\maketitle
\begin{abstract}
The three-state Potts model with antiferromagnetic nearest-neighbor (n.n.)
 and ferromagnetic next-nearest-neighbor (n.n.n) interaction
is investigated within a mean-field theory.
We find that the phase-diagram contains two kind of ordered phases,
so-called BSS phase and PSS phase, separated by a discontinuous 
phase transition line.
Order-disorder transition is continuous for the weak n.n.n. interaction
and becomes discontinuous transition when the 
n.n.n. interaction is increased.
We show that the multicritical point where the order-disorder transition
becomes discontinuous is indeed a tricritical point.

\end{abstract}
\newpage
\section{INTRODUCTION}
Antiferromagnetic three-state Potts model 
has interesting properties. 
It is described by the following Hamiltonian:
\be
H=J\sum_{<ij>}\delta (s_i,s_j) \label{nn}
\ee
where $<ij>$ indicates summation over nearest neighbor pairs, and
$s_i=1,2,3$ denotes three-state Potts spin on the $i$'th
site.
In antiferromagnetic case ($J>0$),
the neighboring two spins prefer to take different values.
Typical ground state configuration of the model
on the square lattice is depicted in Fig.(1).
One can change the state of certain spin in Fig.(1) from ``2'' to ``3'',
or vice versa, without any energy cost.
One half of all the spins are such
``semi-free'' spins, therefore ground state is infinitely degenerated.
It should be noted that if we divide the lattice in Fig.(1) into
two interpenetrating sublattices, there are only ``1'' spins 
on one sublattice, 
while random mixture of ``2'' spins and ``3'' spins is present
on another sublattice.

For the model on simple cubic lattice, it is known that
long-range order is realized
at finite temperature \cite{Wang89,Wang90}
in spite of the high degeneracy of ground-state
which usually leads to the disordered state 
at all temperature range \cite{Wannier50}.
Recently this model (on the simple cubic lattice) has been
studied intensively concerning two interests;
one on the nature of the order-disorder transition 
and another on the
nature of ordered phase.
\par
Order-disorder transition is continuous and the associated critical 
exponents are estimated by various methods, such as Monte Carlo
simulation \cite{Wang89,Banavar80,Ueno89,Ueno93,Gottlob94,Gottlob942}
and the coherent-anomaly method \cite{Kolesik95}.

As for the nature of ordered phase, it is known that
so-called broken-sublattice-symmetry (BSS) phase is realized
at a sufficiently low temperature \cite{Banavar80,Okabe92},
in which one of the sublattices is dominated by one spin state,
while the second sublattice is dominated by the mixture of the remaining
two spin states.

Recently several different
claims have been made
about the nature of the ordered phase at the 
region just below the transition temperature.
Lapinskas and Rosengren concluded that 
so-called permutationally-symmetric-sublattice (PSS) phase,
in which each two sublattices are dominated by one spin state, 
is realized at a very narrow temperature range
just below the transition point, based on cluster-variation method
analysis 
\cite{Lapinskas94,Rosengren93}
 and Monte Carlo simulation \cite{Kundrotas95}.
On the other hand, Kolesik and Suzuki have performed Monte Carlo simulations
and observed
a rotationally symmetric phase 
at a certain range near the transition point, finding no evidence of
the PSS phase \cite{Kolesik952}.

Now we consider the effect of next-nearest-neighbor (n.n.n.)
ferromagnetic interaction.
Let us consider the following Hamiltonian:
\be
H=J \sum _{<i,j>_{nn}} \delta(s_i,s_j)- \gamma J \sum_{<i,j>_{nnn}}
\delta(s_i,s_j)
\label{hamil}
\ee
where the first and the second summation runs over
all nearest neighbor and next-nearest neighbor pairs on the
simple cubic lattice, respectively.
We assume that $J>0$ and $\gamma \geq 0$.
The lattice consists of two sublattices, which
we refer as A and B.
The first term in (\ref{hamil}) represents 
the inter-sublattice antiferromagnetic
interaction and the second one represents 
the intra-sublattice ferromagnetic interaction,
which resolves the high degeneracy of ground state.\par
Thus the n.n.n. interaction affects the nature
of the ordered phase because the BSS phase
costs energy proportional to $\gamma J$ compared to the PSS phase.
This effect for the models on the square
lattice has been studied by several methods
\cite{Grest81,Nijs82,Ono84}.
\par

In three dimensions, we expect another effect
produced by a strong n.n.n. interaction; 
one can consider $\gamma \rightarrow \infty$ limit
as two independent systems of the ferromagnetic $q=3$ Potts model, which 
undergoes the first-order phase transition 
in three dimensions \cite{Fukugita89}.
Thus the order-disorder transition is expected to become
discontinuous as $\gamma$ becomes large.\par

Mean-field theory gives qualitatively 
correct answer for both $\gamma=0$(AF) and 
$\gamma \rightarrow \infty$ (F)
cases in three or greater dimensions,
namely the BSS phase is realized below the continuous
phase transition point in the AF case, and discontinuous phase transition
occurs in the F case.
We investigate the intermediate region 
using mean-field theory in the following sections.
\par

\section{MEAN-FIELD ANALYSIS}
To perform a mean-field calculation,
we use a method of variational free energy which is equivalent to
equation of self-consistency \cite{Binney}.
Let us consider the following mean-field Hamiltonian:
\be
H_0=\sum_i \sum_{s=1}^3 h_{is}\delta(s_i,s) \label{h0}
\ee
where $h_{is}$ is a mean-field acting on the $i$'th spin.
Boltzmann probability factor for the system described by $H_0$
is denoted by $P_0(\{s_i\})$, which is a product of 
probability factors of individual spins:
\be
P_0(\{s_i\})=\Pi_i p_i(s_i) \label{p0}
\ee
\be
p_i(s)={\exp(-\beta h_{is})
\over 
\exp(-\beta h_{i1})+
\exp(-\beta h_{i2})+
\exp(-\beta h_{i3})
}
\label {pi}
\ee
We minimize the following variational free energy
(divided by the total number of spins $N$)
with respect to $h_{is}$:
\be
F_0=(<H>_0-T S_0)/N  \label{f-0}
\ee
The symbols
$<\cdots>_0$ and $S_0$ denotes respectively
the expectation value and the entropy 
associated with the probability distribution $P_0(\{s_i\})$:
\begin{eqnarray}
<H>_0&=&J \sum _{<i,j>_{nn}} \sum_s p_i(s)p_j(s)
-\gamma J \sum_{<i,j>_{nnn}} \sum_s p_i(s)p_j(s) 
\label{e-0} \\
S_0&=&-\sum_i \sum_s p_i(s)\ln p_i(s)
\label{s-0}
\end{eqnarray}
Minimization problem with respect to $h_{is}$
is equivalent to that with respect to $p_i(s)$ under following
constraints:
\be
 p_i(1)+p_i(2)+p_i(3)=1 
\,\,\,\,\,
, 
\,\,\,\,\,
 p_i(s)\geq 0 
\ee
Translational symmetry assures that the free energy (\ref{f-0})
 is minimized when:
\begin{eqnarray}
p_i(s)&=&C_{sA} \,\, \mbox{for all} \,\, i\in A \\
p_i(s)&=&C_{sB} \,\, \mbox{for all} \,\, i\in B
\end{eqnarray}
Obviously, $C_{s\alpha}$ coincides with the expectation value of
the concentration of $s$'th spin state on the 
sublattice $\alpha$, calculated with the probability distribution $P_0\{s_i\}$.
They are under following constraints:
\be
C_{1\alpha}+C_{2\alpha}+C_{3\alpha}=1 
\,\,\,\,\,
, 
\,\,\,\,\,
C_{s\alpha}\geq 0  \label{constr}
\ee
Then $<H>_0$ and $S_0$ are expressed with $C_{s\alpha}$:
\be
<H>_0={ z_1 J N \over 2}\sum_s C_{sA}C_{sB}
-{z_2 \gamma J N\over 2}\sum_s (C_{sA}^2+C_{sB}^2)/2
\ee
\be
S_0
=-N\sum_s (C_{sA}\ln C_{sA}+C_{sB}\ln C_{sB})/2
\ee
where $z_1 ,z_2$ denotes the coordination number of n.n. and n.n.n.
sites respectively (for the simple cubic lattice, $z_1=6$ and $z_2=12$).

Furthermore, we define two-component 
sublattice magnetization similar as Ono used \cite{Ono86}:
\begin{eqnarray}
x_\alpha&=&(C_{1\alpha}-C_{2\alpha})/\sqrt{3} \\
y_\alpha&=&(C_{1\alpha}+C_{2\alpha}-2C_{3\alpha})/3
\end{eqnarray}
Then the three quantities $C_{1\alpha}$,$C_{2\alpha}$, and $C_{3\alpha}$
 can be expressed by two quantities $x_\alpha$ and $y_\alpha$
owing to the constraint $C_{1\alpha}+C_{2\alpha}+C_{3\alpha}=1$. 
\begin{eqnarray}
C_{1\alpha}&=&{1\over 3}+{\sqrt{3}\over 2} x_\alpha +{1 \over 2} y_\alpha \\
C_{2\alpha}&=&{1\over 3}-{\sqrt{3}\over 2} x_\alpha +{1 \over 2} y_\alpha \\
C_{3\alpha}&=&{1\over 3}-y_\alpha
\end{eqnarray}
The two-component sublattice magnetization $(x_\alpha,y_\alpha)$
carry an irreducible, unitary representation of the permutation group of 
the three spin states $\{1,2,3\}$.
Owing to the constraints $C_{s\alpha}\geq 0$,
the sublattice magnetization
$(x_\alpha,y_\alpha)$ is restricted to take a value 
within a regular triangle
on the $x_\alpha-y_\alpha$ plane (Fig.2). 
Three vertexes of the triangle correspond to
completely ordered state
($C_{1\alpha}=1$,$C_{2\alpha}=0$,$C_{3\alpha}=0$ etc.)
and the point $(0,0)$ corresponds to completely disordered state
($C_{1\alpha}=C_{2\alpha}=C_{3\alpha}=1/3$).

We have minimized the free energy $F_0$
with respect to $x_A$,$y_A$,$x_B$, and $y_B$
numerically using the following iteration:
\begin{eqnarray}
X_{\alpha}^{(n+1)} & = & 
X_{\alpha}^{(n)} - \Delta {\partial F_0 \over \partial x_\alpha}
|_{x_\alpha=X_\alpha^{(n)},y_\alpha=Y_\alpha^{(n)}}
\nonumber  \\
Y_{\alpha}^{(n+1)} & = & 
Y_{\alpha}^{(n)} - \Delta {\partial F_0 \over \partial y_{\alpha}} 
|_{x_\alpha=X_\alpha^{(n)},y_\alpha=Y_\alpha^{(n)}}
\label{num}
\end{eqnarray}
where $\Delta$ is a small, positive quantity.
The iteration (\ref{num}) is repeated until $F_0$ converge
to some minima.
Several values were used as an initial value 
$X_\alpha^{(0)}$ and $Y_\alpha^{(0)}$ to find
absolute minimum value of $F_0$ out of all minima.
\par
Finally we obtained a phase diagram 
of two parameters $T/J$ and $\gamma$ (Fig.(3)).
The order-disorder transition is continuous for $\gamma \leq 3/2$
and discontinuous for $\gamma >3/2$.
There are two kind of ordered phases, BSS phase and
PSS phase, separated by a discontinuous transition line.
Typical values of the concentrations of the three states 
and the sublattice magnetizations
of each ordered phases are shown in Fig.(4).
In the PSS phase, the angle between each sublattice magnetizations
is non-trivial value; it is greater than $120^\circ$ and less than
$180^\circ$.
The non-trivial value of the angle
can be understood as follows.
The antiferromagnetic n.n. interaction makes two points
$(x_A,y_A)$ and $(x_B,y_B)$ ``repulsive'',
therefore prefers $180^\circ$.
On the other hand, the energy of the ferromagnetic n.n.n. interaction
is minimized when
$(x_\alpha,y_\alpha)$ locates on the vertex of the triangle,
therefore prefers $120^\circ$.
Thus we can understand the non-trivial value of the angle
as a result of the competition of the two interactions.

Banavar and Wu have studied the same model as (\ref{hamil})
with $q=3,4$ using 
the mean-field theory \cite{Banavar84} 
and concluded that
the PSS phase does not appear.
Our result disagree with theirs.
\section{EFFECTIVE FREE ENERGY FORM}
Since the numerical method (\ref{num})
becomes less precise near the  
critical line, we expand the mean-field free energy $F_0$
in powers of the order-parameter
to investigate the critical behavior.\par
Firstly we define ferromagnetic and antiferromagnetic order-parameters:
\begin{eqnarray}
\xf=x_A + x_B \,\,,\,\, \xaf=x_A - x_B \nonumber \\
\yf=y_A + y_B \,\,,\,\, \yaf=y_A - y_B
\end{eqnarray}
Relevant quantities to the order-disorder transition are
$x_{AF}$ and $y_{AF}$, which carry 
an irreducible representation of
a group $P_3$ (permutation of three spin states) $\times P_2$
(permutation of two sublattices), isomorphic to the group $c_{6v}$ 
(point group of a regular hexagon).
Indeed, the allowed range of the antiferromagnetic order-parameter
$(x_{AF},y_{AF})$ is a regular hexagon in the $x_{AF}-y_{AF}$ plane.

Then we introduce polar coordinates
of ferromagnetic and antiferromagnetic order-parameter:
\begin{eqnarray}
\xf = R_F    \cos \theta_F    &\,\,,\,\, &  \yf = R_F    \sin \theta_F   \nonumber \\
\xaf = R_{AF} \cos \theta_{AF} &\,\,,\,\, &  \yf = R_{AF} \sin \theta_{AF}
\label{polar} 
\end{eqnarray}
The PSS phase and the BSS phase can be distinguished by the direction of
the antiferromagnetic order-parameter $\theta_{AF}$; the value of 
$\theta_{AF}$
expected in the PSS and the BSS phase is $k\pi/3$ and
$(k+1/2)\pi/3$ ($k=0,1,2,3,4,5$), respectively.
Thus a quantity $\cos 6\theta_{AF}$ is a relevant
quantity to the PSS-BSS phase transition \cite{Kolesik952}.
The value of $\cos 6\theta_{AF}$ is $-1$ in the BSS phase,
while $\cos 6\theta_{AF}=1$ in the PSS phase.

Now we use $R_{AF}$,$\theta_{AF}$,$R_F$, and
$\theta_F$ as independent variables of the free energy $F_0$
and
trace out $R_F$ and $\theta_F$
in order to obtain an effective free energy form 
which is expressed by $R_{AF}$ and $\theta_{AF}$ only.
\be
F_{AF}(R_{AF},\theta_{AF})\equiv
\min_{R_F,\theta_F}F_0(R_{AF},\theta_{AF},R_F,\theta_F)
=F_0(R_{AF},\theta_{AF},\tilde{R_F},\tilde{\theta_F})
\label{faf}
\ee
where $\tilde{R_F}$ and $\tilde{\theta_F}$ gives minimum value of $F_0$ for
fixed $R_{AF}$ and $\theta_{AF}$.
Location of minima 
$\tilde{R_F}$ and $\tilde{\theta_F}$ 
is determined by solving the following equations: 
\be
{\partial F_0 \over \partial R_F}|_
{(R_F,\theta_F)=(\trf,\ttf)}=0
\label{min1}
\ee
\be
{\partial F_0 \over \partial \theta_F}|_{
(R_F,\theta_F)=(\trf,\ttf)}=0
\label{min2}
\ee
Since we cannot solve (\ref{min1}) and (\ref{min2}) explicitly,
we expand $\tilde{R_F}$ and $\tilde{\theta_F}$
in powers of $R_{AF}$ which is small around the critical line.
Firstly we expand (\ref{min1}) and (\ref{min2})
in powers of both $R_F$ and $R_{AF}$.
Lowest order terms follow:

\begin{eqnarray}
{\partial F_0 \over \partial R_F} &=& 
{9(2J-4 \gamma J+T)\over 8}R_F 
-{27 T \sin(2 \theta_{AF}+\theta_F)\over 64}
R_{AF}^2   \nonumber \\ 
& & +O(R_F^2)+O(R_{AF}^4) +O(R_F R_{AF}^2)  \label{exp11}\\
\nonumber \\
{\partial F_0 \over \partial \theta_F} &=&
R_F\left(
-{27 T \cos(2 \theta_{AF}+\theta_F)\over 64}
R_{AF}^2 \right)\nonumber \\
& & +R_F \left(O(R_{AF}^4)+O(R_F^2)+O(R_{AF}^2 R_F)
\right) \label{exp12}
\end{eqnarray}

Equations (\ref{exp11}) and (\ref{exp12}) indicate that
 $\trf\sim O(R_{AF}^2)$ and
$\cos(2 \theta_{AF}+\ttf)\sim O(R_{AF}^2)$, so
we assume that $\trf$ and $\ttf$ can be expanded as below:
\begin{eqnarray}
\trf &=& c_2 R_{AF}^2 + c_4 R_{AF}^4 + c_6 R_{AF}^6 \cdots 
\label{exp21}\\ 
\ttf &=& \pi/2 -2\theta_{AF} + d_2 R_{AF}^2 + d_4 R_{AF}^4 + d_6 R_{AF}^6
\cdots
\label{exp22} 
\end{eqnarray}
The coefficients can be determined by letting 
(\ref{exp21}) and (\ref{exp22}) into 
equations (\ref{min1}) and (\ref{min2}), leading to the following result: 
\begin{eqnarray}
c_2 =& 3\para/8 \\
d_2 =& {9\over 64}(-2+4\para-\para^2)\sin(6\theta_{AF}) \\
c_4 =& {27\over 64}(\para-\para^2)+{27\over 512}
(-2\para +4\para^2 -\para^3)\cos(6\theta_{AF}) \\
d_4 =& {81\over 256}(-1+4\para-4\para^2+\para^3)\sin(6\theta_{AF})
\nonumber\\
&+{-81\over 8192}(2-4\para+\para^2)^2 \sin(12\theta_{AF}) \\ 
\vdots \nonumber
\end{eqnarray}
where $t=T/J$ and $\para=t/(2-4\gamma+t)$.

Finally the effective free energy form is obtained by letting
(\ref{exp21}) and (\ref{exp22})
into (\ref{faf}).
Note that only the terms
which is invariant under the transformations of $c_{6v}$
are present in $F_{AF}$, and they can be written as 
$R_{AF}^{6n +2m}\cos(6n \theta_{AF})\,\,,(n,m\geq 0)$. 
We have calculated the effective free energy $F_{AF}$ 
up to sixth order term as below:
\be
F_{AF}=A_0 + A_2 R_{AF}^2 + A_4 R_{AF}^4 
+(A_6 + B_6 \cos 6\theta_{AF})R_{AF}^6
+O(R_{AF}^8)
\ee
where
\begin{eqnarray}
A_0=&J - 2 \gamma J - T \log 3 \\
A_2=&{9\over 16}J(-2 - 4 \gamma + t) \\
A_4=&{81 \over 1024}T(2-\para) \\
A_6=&{243 \over 8192}T(4-6\para+3\para^2) \\
B_6=&{243 \over 163840}T(-8+30\para -30\para^2+5\para^3) 
\end{eqnarray}

The critical line is the region $A_2=0$ and $A_4 \geq 0$,
which corresponds to
the region $t=2+4\gamma \,\,,\,\,\gamma\leq 3/2$. 
The order-disorder transition is discontinuous for $\gamma >3/2$
and $(\gamma,t)=(3/2,8)$ is a tricritical point where $A_2$ and $A_4$ vanish
simultaneously \cite{TCP}.

The six-fold anisotropy term $B_6R_{AF}^6\cos6\theta_{AF}$
is the origin of the BSS and the PSS phase; positive
$B_6$ corresponds to the BSS phase and negative $B_6$ to the PSS phase.
Presence of higher order anisotropic
terms such as $R_{AF}^{12}\cos 12\theta_{AF}$
allows a occurrence of minima of $F_{AF}$ 
at $\theta_{AF}\neq k\pi/6$, which corresponds to neither
the PSS nor the BSS phase.
However, the positions of minima remain at $\theta_{AF}=k\pi/6$ if 
the coefficients of such higher order anisotropic terms are
sufficiently small.
The following example helps understanding above issue:
a positions of minima and maxima of a function
$f(\theta)=\cos \theta +a \cos 2\theta$ remain at
$\theta=k\pi$, as long as $|a|\leq 1/4$.

On the critical line $t=2+4\gamma$, $B_6$ is expressed as a function of
$\gamma$:
\be
B_6=243(2+4 \gamma)(1+30 \gamma-180 \gamma^2+40 \gamma^3)/1310720
\ee
It vanishes at $\gamma \sim 0.2032$ on the critical line,
where the BSS-PSS transition line merges into the critical line.

Note that higher order term is needed to investigate off-critical regions,
where the numerical method (\ref{num}) works well. So we do not calculate 
higher order terms.

\section{CONCLUSION}
In this paper we have shown that a mean-field theory analysis 
indicates the presence of a tricritical point in
the three-state Potts model 
with the antiferromagnetic n.n. interaction and 
the ferromagnetic n.n.n. one.
It is believed that mean-field theory is qualitatively correct
above the upper critical dimension, where
spatial fluctuations of
order-parameter at the critical region become negligible.
Since the upper critical dimension of the tricritical phenomena
is three \cite{TCP}, we expect the model on the simple cubic lattice
to posses a tricritical point.
Owing to the simpleness of the Hamiltonian, this model must  serve to
develop numerical methods for studies of a tricritical phenomena,
such as finite-size scaling in Monte Carlo simulations.

As for the nature of the ordered phase,
we have shown that the PSS phase is realized at 
the strong n.n.n. coupling region, as a result of the
competition of the two kind of interactions,
while the BSS phase is realized at the weak n.n.n. coupling region.

However, it should be noted that 
mean-field type analysis like this work may not give
collect information, as pointed in Ref.\cite{Kolesik952},
 about the six-fold anisotropy term
which becomes relatively small compared to the order-parameter 
fluctuation near the critical line.
So further study, such as Monte Carlo simulations, may be needed 
to clarify the nature of the ordered phase just below the
critical line.

\begin{center}
{\bf Acknowledgment}
\end{center}
The author is indebted to S.~Hikami for reading the manuscript.
This work was supported by a Grant-in-Aid for Scientific Research
by the Ministry of Education, Science and Culture.
\newcommand{\cit}[4]{{#1} {\bf #2}, #3 (#4).}
\def\prl{Phys. Rev. Lett.}
\def\pr{Phys. Rev.}
\def\jpsj{J. Phys. Soc. Jpn.}

{\bf Figure Captions }

Fig.1.
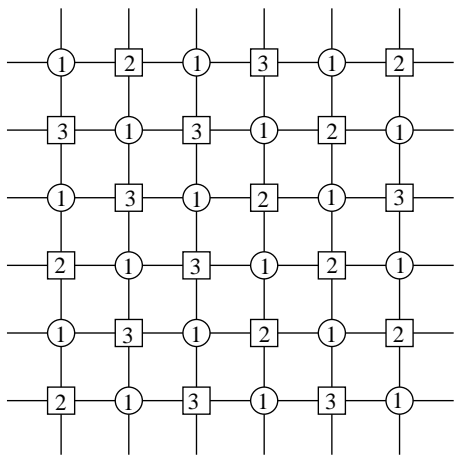
Typical ground state configuration of antiferromagnetic
three-state Potts model on a square lattice.
Two different symbols (
\begin{picture}(6,6) \put(0,0){$\Box$}\end{picture}
and
\begin{picture}(6,6) \put(3,3){\circle{6}}
 \end{picture}) show the two interpenetrating sublattices.

Fig.2.
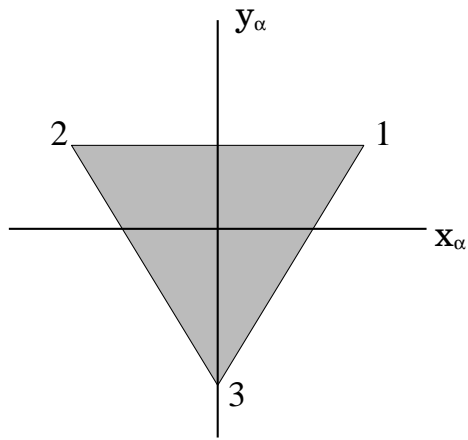
A range of sublattice magnetization $(x_\alpha ,y_\alpha)$.
Three vertexes 1,2, and 3 correspond to the completely ordered states
in which all the spins in the sublattice take the same value
1,2, and 3, respectively.

Fig.3.
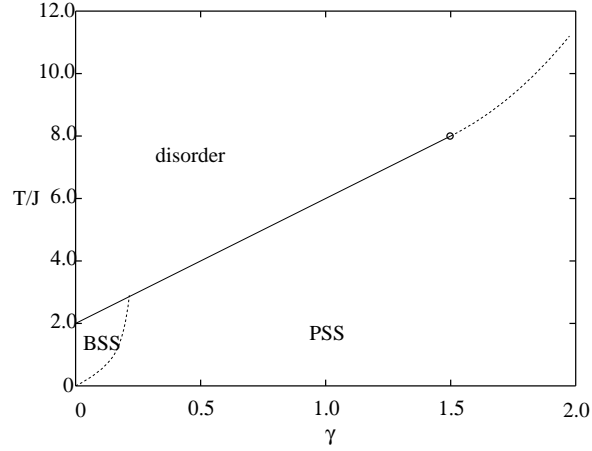
Schematic phase diagram of two parameters $T/J$ and $\gamma$.
The solid line indicates a continuous transition and the dashed line
indicates a discontinuous one.
The continuous order-disorder transition line ends at $\gamma=1.5$.
The BSS--PSS phase boundary merges into the order-disorder transition line
at $\gamma\sim 0.2032$.

Fig.4.
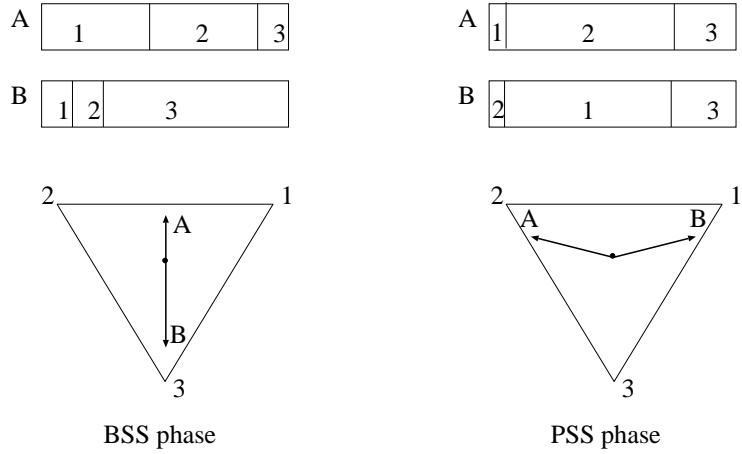
Typical values of the concentrations of 
the three states (upper) and 
the sublattice magnetizations (lower) 
of each ordered phases.

\end{document}